# Superconductivity in twisted bismuth bilayers

*Isaías Rodríguez, Renela M. Valladares, David Hinojosa-Romero, Alexander Valladares, Ariel A. Valladares\**

I. Rodríguez, D. Hinojosa-Romero, A. A. Valladares

Instituto de Investigaciones en Materiales, Universidad Nacional Autónoma de México, Apartado Postal 70-360, Ciudad Universitaria, CDMX, 04510, México

*E-mail: valladar@unam.mx

R. M. Valladares, A. Valladares

Facultad de Ciencias, Universidad Nacional Autónoma de México, Apartado Postal 70-542, Ciudad Universitaria, CDMX, 04510, México.

Two twisted bismuth bilayers, TBB, each with 120 atoms, are studied by means of their electronic density of states and their vibrational density of states using first principles calculations. Metallic character at the Fermi level is found for the non-rotated sample as well as for each sample rotated 0.5°, 1.0°, 1.5°, 2.0°, 2.5°, 3.0°, 4.0°, 5.0°, 6.0°, 7.0°, 8.0° and 10° with respect to each other. Assuming that the superconductivity is BCS-type and the invariance of the Cooper pairing potential, we predict a maximum superconducting temperature $T_C$ ~ 1.8 K for a magic angle of 0.5 degrees between the two bilayers, increasing the superconducting transition temperature from the experimentally measured value of 0.53 mK for the Wyckoff structure of crystalline bismuth.

## 1. Introduction

For more than 100 years superconductivity has been present in scientific research. Sometimes in the forefront, sometimes in the background, but the hope of finding high temperature superconductors, the holy grail of technology, has prevail over a century. This discovery would certainly revolutionize several fields of knowledge and technology. Up to the present, disappointment has been the result of multiple claims of having found superconductivity at room temperature. Investigations of materials at high pressures have given hope, once more, towards this discovery, however some of these claims are either for very high pressures,[1–4] or for new materials that have not been reproduced worldwide.[5–8] High pressure experiments are very difficult to conduct since the pressures needed, of the order of giga pascals, can only be applied for very short times. One should ask whether pressures below atmospheric may be a different and promising way to approach the problem.[9–12] On the other hand, the new materials so far reported have to be validated by several laboratories to accomplish credibility. Some known materials, like the Xenes,[13,14] have also been the subject of study in this quest.

Xenes are a class of two-dimensional (2D) single-element layered materials, primarily derived from group IV elements like carbon (graphene),[15] group V elements like phosphorus (phosphorene),[16] or group VI elements like tellurium (tellurene).[17] These materials demonstrated unique physical and chemical characteristics against the corresponding three-dimensional (3D) structures. Xenes have sparked interest due to their tunable properties and potential applications in electronics and photonics.[18] Xenes electronic and lattice dynamics properties can be engineered by various techniques, like adjusting the lattice structure, doping the layers, twisting the layers, etc.

Bismuth xenes, also known as bismuthenes, have several 2D layered allotropes that have been predicted theoretically,[19,20] and at least two of these phases have been experimentally confirmed.[21,22] The first one is an α- phase also known as Bi(110) and consist of puckered orthorhombic structures similar to black phosphorous, while the β- phase also known as Bi(111) consists of a buckled honeycomb-like structure similar to silicene.[20,21,23]

Bismuthene, as several other xenes, have been the subject of much interest and investigation as a topological insulator.[24–27] However, the discovery of superconductivity in twisted bilayer graphene (TBG)[28,29] opens up a new field of study for bismuthene.

Bismuth superconductivity was discovered in amorphous bismuth ($a$-Bi)[30–33] with a critical transition temperature, $T_c$, of ~ 6 K, while superconductivity in crystalline bismuth ($x$-Bi) was for the first time predicted to have a $T_c$ lower than 1.5 mK,[34] and later measured with a $T_c$ of 0.53 mK.[35] Also, the other stable crystalline phases of bismuth at high pressure were studied finding a maximum of $T_c$ ~ 8 K for the Bi-V phase.[36,37] In bismuth bilayers a possible $T_c$ of ~ 2.6 K was calculated.[38]

Here we report the possible superconductivity in twisted bismuth bilayers (TBB) of two bilayers of about 5.1 nm long and 4.2 nm wide, that were twisted with respect to one another at angles of 0.5°, 1°, 1.5°, 2°, 2.5°, 3°, 4°, 5°, 6°, 7°, 8°, 9° and 10° forming different Moiré patterns.

## 2. Method

The study of 2D materials forming Moiré patterns has proven to be difficult because a periodic structure that correctly represents the periodicity of the Moiré patterns normally produces large lattice vectors with several thousands of atoms in the unit cell. Several approaches have been considered to circumvent this limitation, such as using classic methods,[39–41] using the tight-binding approximation,[41–44] limiting the structure to particular angles with a manageable number of atoms,[45] or treating rotation as small perturbation over two non-rotated perfect graphene sheets.[46] Here we present a state-of-the-art *ab initio* approach to TBB by constructing a 240 atoms non-periodic nanostructure composed of two 120-atom bismuthene finite sheets.

The nanostructure was constructed starting from the Bi-I (Wyckoff) structure and multiplying it by 10x8x2 to obtain a 960-atom structure (see **Figure 1** A). Then atoms were removed until we ended with two 120-atom bilayers for a total of 240 atoms, (Figure 1 B). To avoid possible self-interactions, we removed periodic boundary conditions, thus ending with two flakes of bismuthene conforming a finite nanostructure, (Figure 1 C). Finally, one of the two flakes of bismuthene from the initial nanostructure was rotated around the perpendicular Z axis 0.5, 1.0, 1.5, 2.0, 2.5, 3.0, 4.0, 5.0, 6.0, 7.0, 8.0, 9.0, and 10.0 degrees, obtaining 13 different structures, each with a different Moiré pattern, (Figure 1 D). The geometry optimization of TBG structures has been known to crinkle the borders of the flake, and to produce creases in the surface, impacting heavily its electronic properties.[45]

The electronic density of states ($N(E)$) and vibrational density of states ($F(\omega)$) were then calculated using DMol$^3$,[47] a first principles simulation code included in the Dassault Systèmes Materials Studio Suite.[48] To obtain the $N(E)$, a single-point energy calculation was performed using double numerical plus polarization functions basis set (dnp),[49] the density functional semi-core pseudo-potential (dspp) was used for the core treatment, the Vosko-Wilk-Nusair exchange-correlation functional was considered within the local density approximation (LDA),[50] a fine integration grid, with octupolar angular momentum fitting functions was incorporated. The real-space cutoff radius for the orbitals was set to 6.0 Å. Due to the size of the nanostructure, as well as the removal of periodic boundary conditions the only relevant k-point is Γ.

$F(\omega)$ was calculated with the finite displacement method (frozen-phonon method), using the same electronic parameters as the $N(E)$ calculation, and a finite difference in the displacement of 0.005 Å to obtain the Hessian matrix. By virtue of the frozen-phonon method, no q-points were used.

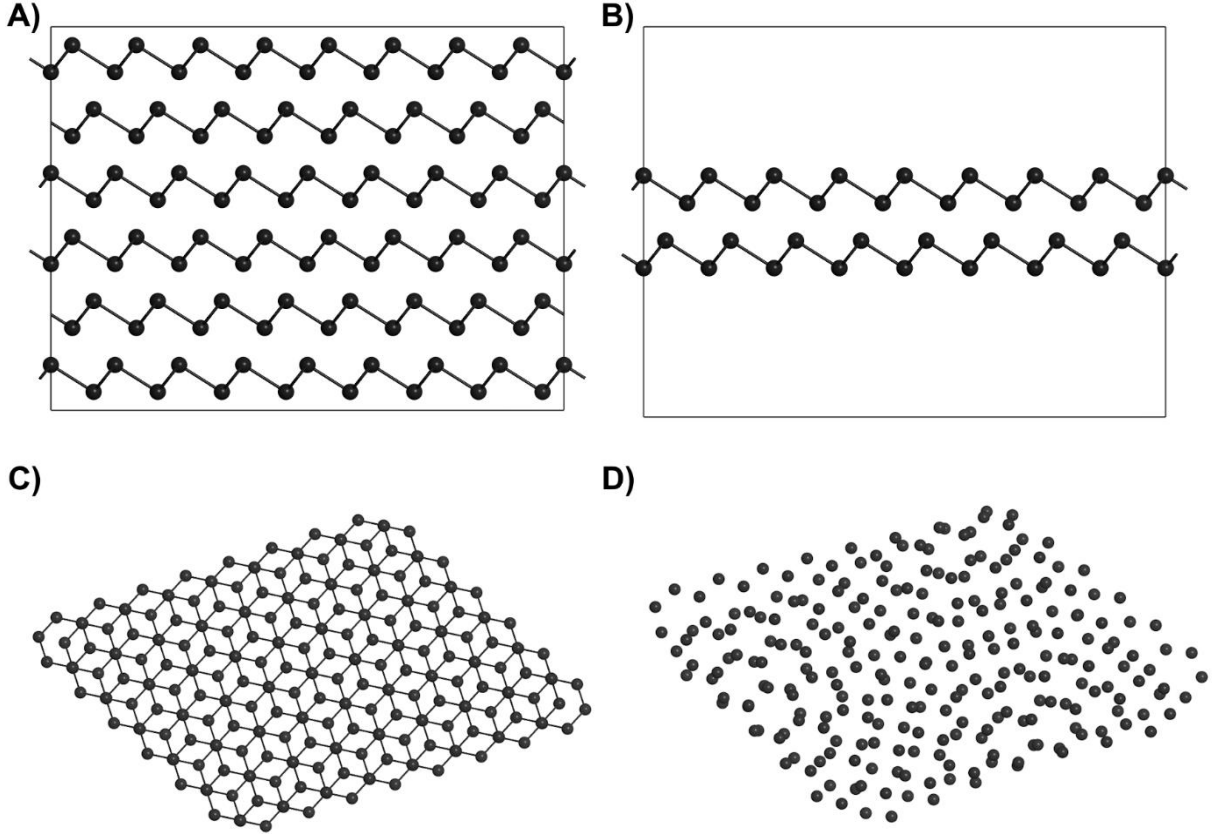

**Figure 1.** Construction of the TBB. A) 10x8x2 supercell of Bi-I (Wyckoff). B) Isolation of two bilayers of 240 atoms. C) Superior view of the two bilayers without periodic boundary conditions. D) 10° rotation of one of the layers with respect to the other; bond-sticks were omitted in D) in order to better appreciate the Moiré pattern.

Then the Debye frequencies and Debye temperatures were calculated using the obtained $F(\omega)$ and the method proposed by Grimvall:[51]

$$\omega_D = \exp\left(\frac{1}{3} + \frac{\int_0^{\omega_{max}} \ln(\omega) F(\omega) d\omega}{\int_0^{\omega_{max}} F(\omega) d\omega}\right), \qquad (1)$$

$$\theta_D = \frac{\hbar \omega_D}{k_B}. \qquad (2)$$

## 3. Results and Discussion

## 3.1. Electronic Density of States

For the *N(E)* determination, the DMol$^3$ analysis tool included in the Materials Studio suite was used, set to eV and an integration method with smearing width of 0.1 eV. The number of points per eV was set to 100. The results were smoothed and analyzed by means of the OriginPro software using a Fast Fourier Transform (FFT) filter of 3-step.[52] The results are given in states per eV per atom in **Figure 2**.

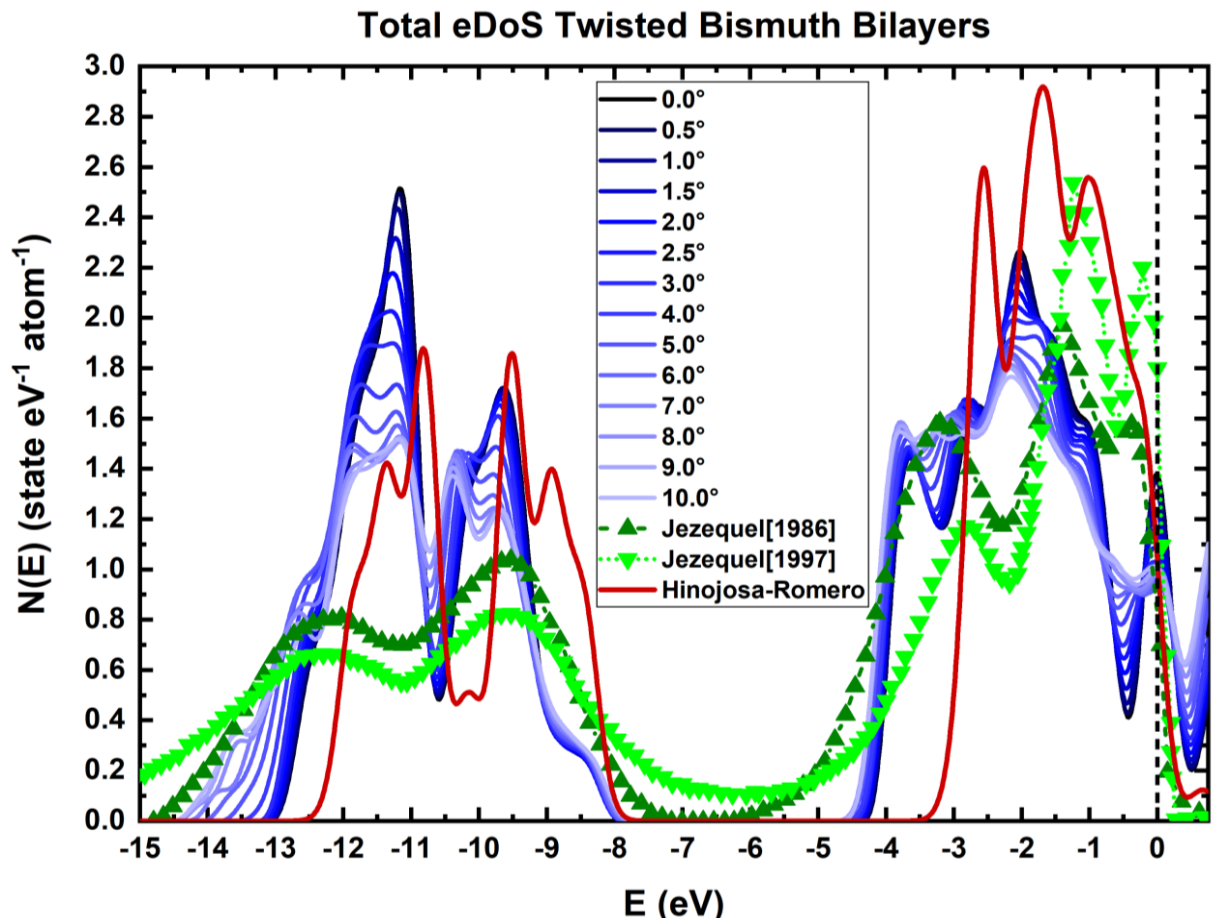

**Figure 2.** Comparison of the electronic density of states (*N(E)*) for twisted bismuth bilayers (TBB). Fermi Level is shown as the dashed vertical line. The electronic density of states is normalized per atom. Experimental results from Jezequel (1986)[53] and Jezequel (1997)[54] are shown in green triangles. DFT calculations from Hinojosa- Romero are represented in red.[38]

The studied samples have a sharp peak on the *N(E)* at the Fermi level (Figure 2), these sudden increases in the *N(E)* are also present in TBG,[55–57] as well as in Moiré nanolayers,[26] and it seems to be caused by an almost flat band near the Γ point. However, the presence of this sharp peak near the Fermi level has not been detected, either in the Bi crystalline structures,[34,37] or in the bismuth bilayers.[26,38,58–60] The peak at Fermi level seems to be caused by the local density of states of the border atoms. The *N(E)* is composed of two main bands; the first band is bimodal and goes from around -15 eV to -8 eV; this first band for all angle rotations between flakes on average includes 43.88±0.07% of s orbital electrons, 34.04±0.08% of p orbital electrons, and 22.08±0.02% of d orbital electrons, while the second band is located from around -4.7 eV to above the Fermi level; this second band has contributions of 70.81±0.26% of p orbital electrons, 25.48±0.12 of d orbital electrons, and 3.71±0.14% of s orbital electrons, , as can be seen in **Figure 3**.

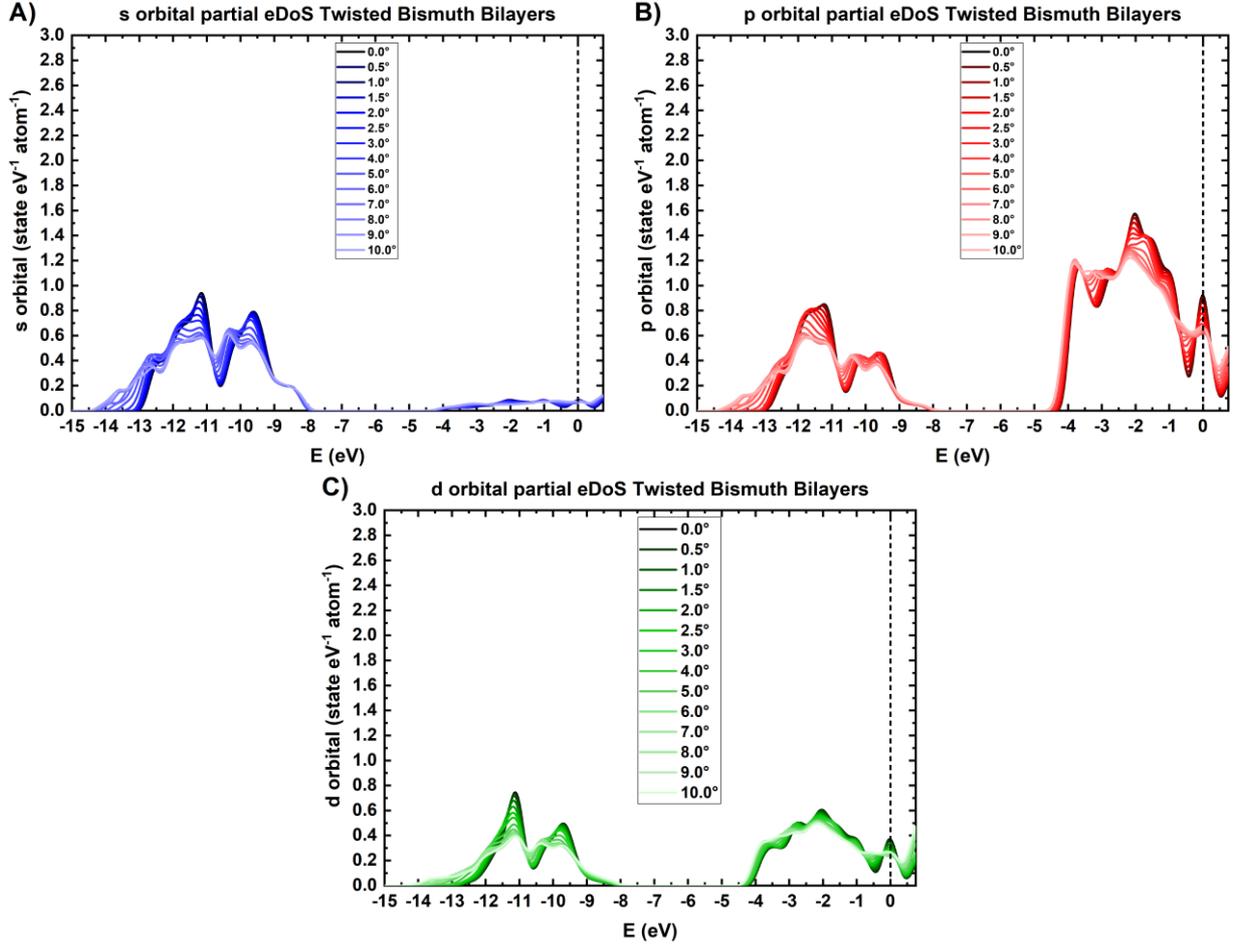

**Figure 3.** Partial electronic density of states for TBB. A) s orbital partial in blue shades, B) p orbitals in red shades, C) d orbitals in green shades, the Fermi level is at the vertical dashed line.

Similarly, the first band contains 91.11±0.32% of the s orbital electrons, with the remaining 8.89±0.32% in the second band, the first band contains 29.41±0.10% of the p orbital electrons, while the second band contains the remaining 70.59±10.38%, finally the first band contains 42.89±0.17%, with the 57.11±0.16% remaining d orbital electrons in the second band as can be seen in **Figure 4**.

For the first band, the s orbital contributions had a maximum of 43.98% for the non-rotated flakes and has a monotonic decreasing behavior to a minimum of 43.78% for the 8° rotation between the flakes, Figure 4 A), and finally increases for the last two studied samples. The p orbital contribution had a minimum of 33.96% for the non-rotated flakes, and has a monotonic increasing behavior to a maximum of 33.16% for the 9° rotation between flakes, and finally decreases for the 10° sample, Figure 4 B). The p orbital contribution starts at 22.06% for the non-rotated flakes, and monotonically increases for the first 8 samples to a maximum of 22.10% for the 4° rotation between flakes, then it monotonically decreases to a minimum of 22.05% for the 10° rotation between flakes, Figure 4 C). For the second band, located from -4.7eV to above Fermi level, the s orbital contribution has a minimum of 3.56% for the non-rotated flakes and monotonically increases to a maximum of 3.90% for the 10° rotation between flakes, Figure 4 D). The p orbital contribution has a maximum of 71.12% for the non-rotated flakes and monotonically decreases to a minimum of 70.48% for the 9° rotation and finally increases for the 10° rotation, Figure 4 E). Finally, the d orbital contribution has a minimum of 25.31% for the non-rotated flakes and monotonically increases to a maximum of 25.63% for the 8° rotation between flakes, and then decreases for the last two rotations.

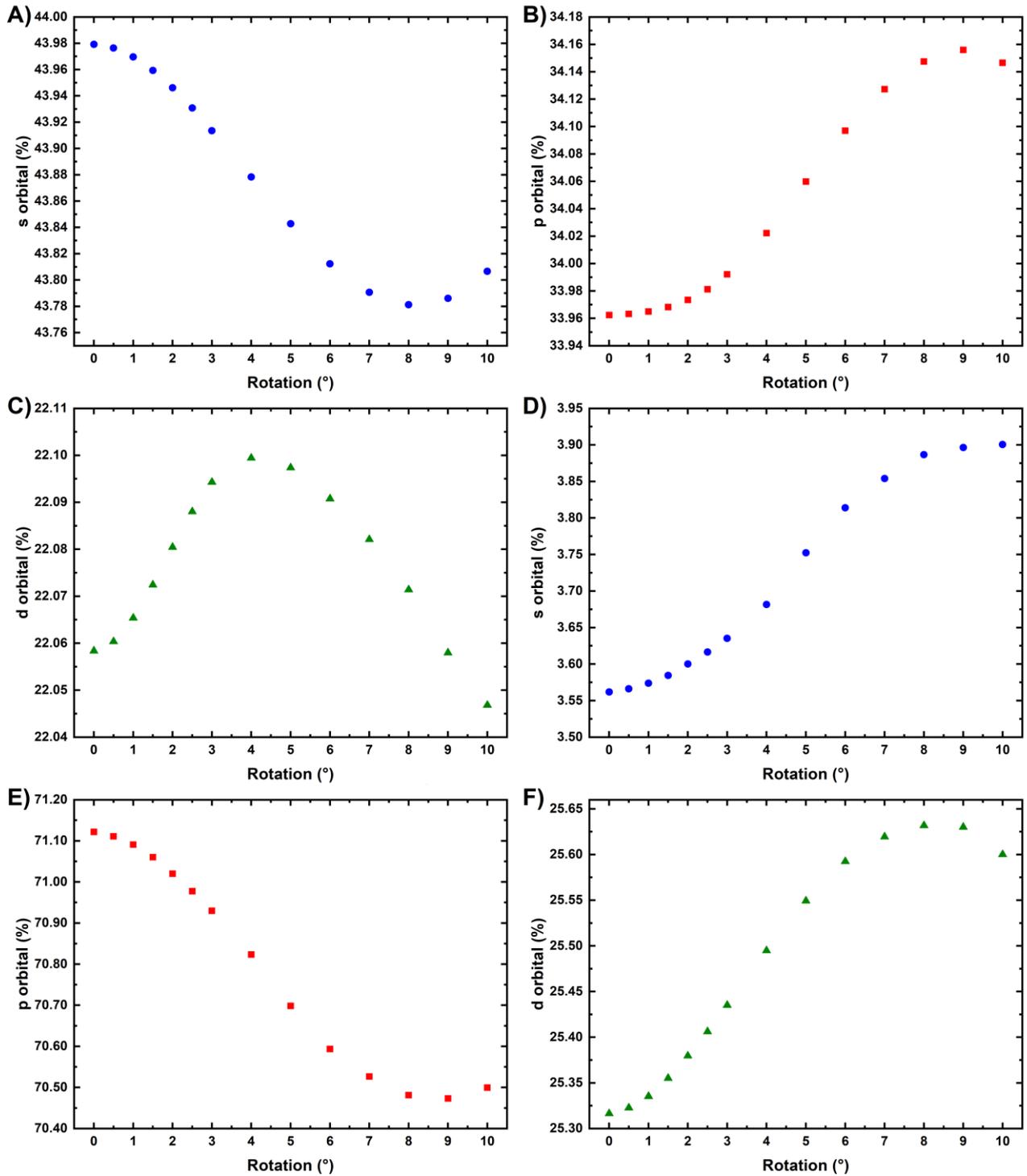

**Figure 4.** Partial orbitals contributions to the bands of TBB. A) s orbital contribution to the first band located between -15eV to -8eV. B) p orbital contribution to the first band. C) d orbital contribution to the first band. D) s orbital contributions to the second band located between -4.7eV to above Fermi level. E) p orbital contribution to the second band. F) d orbital contribution to the second band.

To correctly represent the electronic density of states of a larger bismuth bilayer,[26,38,58–60] as well as the crystalline results,[34,37] it was decided to study the partial electronic density of states of the core (central atoms) of the bilayer disregarding the first 4 atoms from the border on each side of the nanosheets; the partial electronic density of states can be seen in **Figure 5**. The partial eDoS for the core is also represented by two bands, but the sharp peak near the Fermi level is substantially decreased and completely disappears for several of the studied samples (Figure 5), which agrees with all the studies in the literature for bismuth and bismuthene. This seems to indicate that the sharp peak present in the total eDoS is mostly caused by border effects. A more detailed study is currently undergoing and will be reported elsewhere.

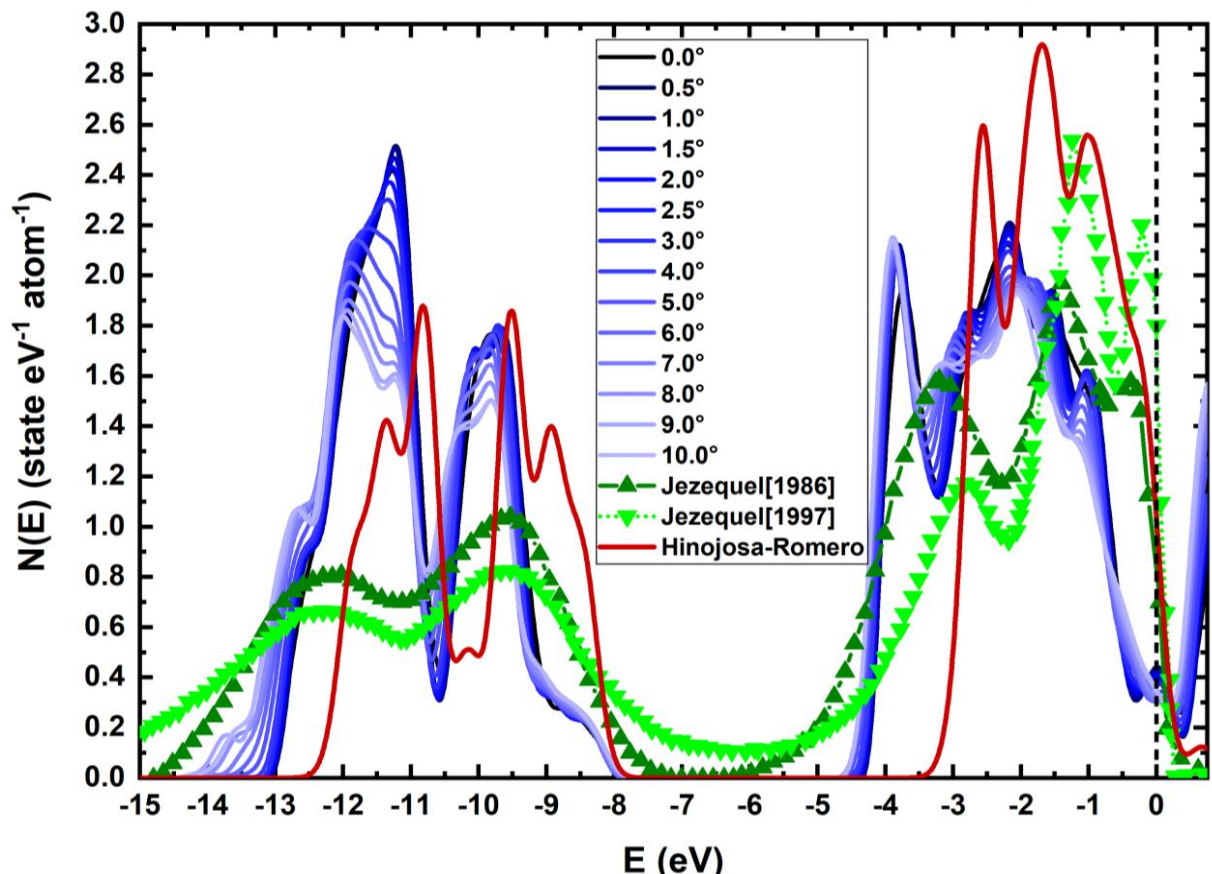

**Figure 5.** Comparison of the partial electronic density of states ($N(E)$) for the core (central atoms) of twisted bismuth bilayers (TBB). Fermi level is at the vertical dashed line. The electronic density of states is normalized per atom in the nanolayers. Experimental results from Jezequel (1986)[53] and Jezequel (1997)[54] are in green triangles. DFT calculations from Hinojosa-Romero are shown in red.[38]

### 3.2 Vibrational Density of States

For the $F(\omega)$ the results were analyzed with OriginPro software,[52] using the normal modes calculated in meV with DMol$^3$. To obtain the $F(\omega)$ a frequency count with a 0.1 meV bin width was used, then the resulting histograms were smoothed with a three-step FFT filter. The translational modes around 0 were removed. Finally, the $F(\omega)$s were normalized to 3, these results can be seen in **Figure 6**.

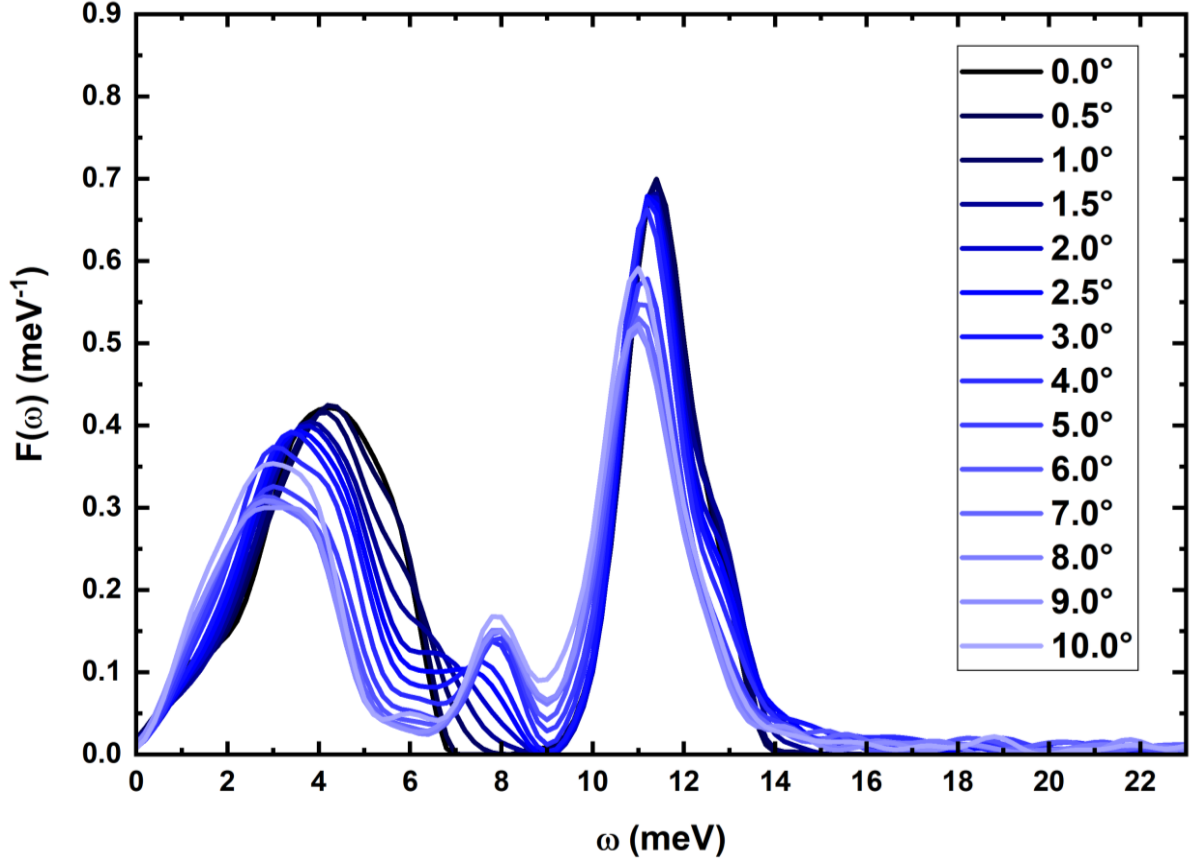

**Figure 6.** Comparison of the vibrational densities of states ($F(\omega)$), normalized to three modes per atom.

The forbidden frequency band-gap from around 7 meV to around 9 meV decreases as the bismuth bilayers are twisted with respect to each other and it completely disappears at the 4.0° rotation, although a minimum of $F(\omega)$ at a frequency of 9 meV is still present for all the studied rotations. The presence of a bimodality in the $F(\omega)$ seems to be a universal characteristic of all 2D materials,[65] and this may be due to the existence of layers as found by our group for 3D materials that have layered structures.

Also, at 2.0° of rotation isolated phonon modes begin to appear above 16 meV, these isolated modes increase in frequency and intensity up to 42 meV for the 10° rotation, although these contributions are small in quantity (less than 0.5% of the modes for the 2.5° sample, and less than 1.8% of the modes for the 10° sample). The relatively high energy of these modes has an impact in related quantities like the Debye frequency and temperature as will be seen in the next section. The existence of these high-energy modes was predicted by Chowdhury,[66] and these modes seem to be related to border effects in the TBB.

### 3.3 Superconductivity

Although there seems not to be an agreement in the kind of superconductivity responsible for the magic angle in TBG,[28,29] there has been a great success using conventional superconductivity for bismuth crystalline phases,[34-37] as well as for bismuth bilayers.[38]

In order to calculate the superconducting transition temperature, the approach developed by Mata-Valladares,[34] is used. The superconducting transition temperature for the crystalline bismuth is calculated with the BCS equation:

$$T_c^\alpha = 1.13\theta_D^\alpha exp\left(-\frac{1}{N(E_F)^\alpha V_0}\right). \tag{3}$$

Analogously, the superconducting transition temperature for the TBB is calculated with the BCS equation:

$$T_c^\beta = 1.13\theta_D^\beta exp\left(-\frac{1}{N(E_F)^\beta V_0}\right), \tag{4}$$

where we have assumed that the electron-phonon coupling potential $V_0$ is the same for both materials, the crystalline and the bilayered structures. If we now take the ratio $T_c^\beta/T_c^\alpha$ and solve for $T_c^\beta$, the following equation is obtained:

$$T_c^\beta = (T_c^\alpha)^{1/\varepsilon}\delta\{1.13\theta_D^\alpha\}^{(\varepsilon-1)/\varepsilon}, \tag{5}$$

where:

$$N(E_F)^\beta = \varepsilon\, N(E_F)^\alpha \text{ and } \theta_D^\beta = \delta\, \theta_D^\alpha.$$

$N(E_F)^\beta$ was calculated directly from Figure 5 and can be seen in **Figure 7** A), while $\theta_D^\beta$ was calculated using Grimvall approach[51], and the results are presented in Figure 7 B). The results considering a frequency cutoff of $\omega_{max} = 23$ meV are shown in red dots, while the results for $\omega_{max} = 50$ meV are shown in black squares. These results offer the exact same $\theta_D^\beta$ up to a twisting angle of 3.0°, and start to differ from 4.0° up to 10.0°. This difference is caused by the high energy isolated modes in the $F(\omega)$. Although these modes represent less than 2% of the total modes calculated, the difference in the Debye temperature can be as large as 6 K for the 9.0° rotation.

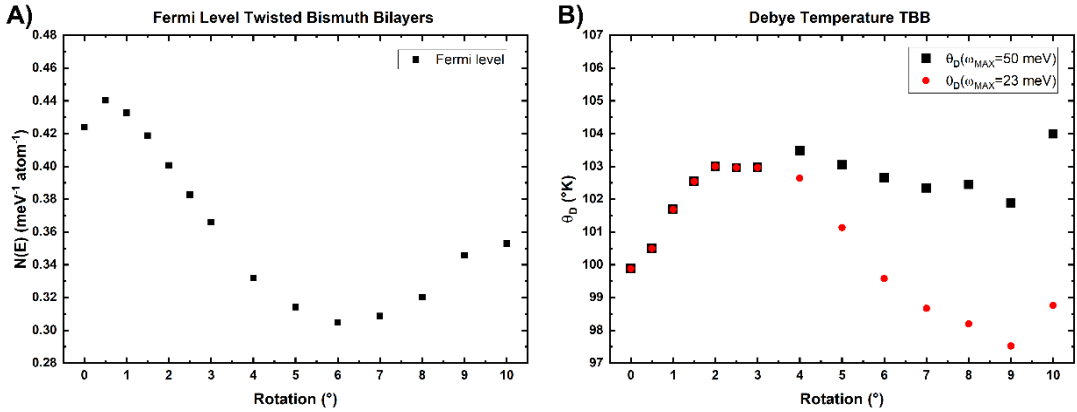

**Figure 7.** A) Electronic density of states at the Fermi level for TBB and B) Debye temperatures for the TBB for two different frequency cut-offs.

Then, using the results found in the literature for crystalline bismuth: $N(E_F)^\alpha = 0.15$ state eV$^{-1}$ atom$^{-1}$ and $\theta_D^\alpha = 134.2$ K[34,37,38] we calculated the superconducting transition temperatures. The results can be seen in **Table 1**, and the corresponding **Figure 8**.

The difference in the superconducting transition temperatures is less than 2.7% whether we consider the high-energy vibrational modes above 23 meV, or not. It also appears that superconductivity is mostly driven by the electronic density of states.

The largest $T_c$ found was 1.85 K for a rotation of 0.5° between layers, as can be seen in Figure 8. A more detailed study should be conducted around this angle to find a possible higher maximum or corroborate if the maximum $T_c$ is indeed 1.85 K. The minimum $T_c$ was 0.35 K for a rotation of about 6.0°. For angles above 7.0° the $T_c$ begins to increase again, so higher rotation angles above 10° may have a higher $T_c$. A more extensive study should be conducted to investigate this surmise.

**Table 1.** Values for the electronic densities of states at Fermi level ($N(E_F)$), Debye temperatures $\theta_D$, the parameters for the Mata-Valladares approach and the superconducting transition temperatures ($T_c$).

| Structure | $N(E_F)$ [state eV$^{-1}$ atom$^{-1}$] | $\theta_D$ [°K] (23 meV) | $\theta_D$ [°K] (50 meV) | $\epsilon$ | $\delta$ (23 meV) | $\delta$ (50 meV) | $T_c$ [°K] (23 meV) | $T_c$ [°K] (50 meV) |
|---|---|---|---|---|---|---|---|---|
| Bi-I | 0.15 | 134.2 | 134.2 | - | - | - | 0.0015 | 0.0015 |
| TBB 0.0° | 0.42 | 99.88 | 99.88 | 2.94 | 0.74 | 0.74 | 1.58 | 1.58 |
| TBB 0.5° | 0.44 | 100.50 | 100.50 | 2.88 | 0.75 | 0.75 | 1.85 | 1.85 |
| TBB 1.0° | 0.43 | 101.69 | 101.69 | 2.79 | 0.76 | 0.76 | 1.76 | 1.76 |
| TBB 1.5° | 0.42 | 102.54 | 102.54 | 2.67 | 0.76 | 0.76 | 1.57 | 1.57 |
| TBB 2.0° | 0.40 | 103.00 | 103.00 | 2.55 | 0.77 | 0.77 | 1.32 | 1.32 |
| TBB 2.5° | 0.38 | 102.96 | 102.96 | 2.44 | 0.77 | 0.77 | 1.08 | 1.08 |
| TBB 3.0° | 0.37 | 102.97 | 102.97 | 2.21 | 0.77 | 0.77 | 0.88 | 0.88 |
| TBB 4.0° | 0.33 | 102.64 | 103.48 | 2.10 | 0.76 | 0.77 | 0.55 | 0.55 |
| TBB 5.0° | 0.31 | 101.14 | 103.04 | 2.03 | 0.75 | 0.77 | 0.41 | 0.41 |
| TBB 6.0° | 0.30 | 99.58 | 102.65 | 2.06 | 0.74 | 0.76 | 0.34 | 0.35 |
| TBB 7.0° | 0.31 | 98.67 | 102.34 | 2.14 | 0.74 | 0.76 | 0.36 | 0.37 |
| TBB 8.0° | 0.32 | 98.20 | 102.45 | 2.31 | 0.73 | 0.77 | 0.44 | 0.45 |
| TBB 9.0° | 0.35 | 97.52 | 101.89 | 2.35 | 0.73 | 0.76 | 0.64 | 0.67 |
| TBB 10.0° | 0.36 | 98.76 | 104.00 | 2.83 | 0.74 | 0.77 | 0.72 | 0.76 |

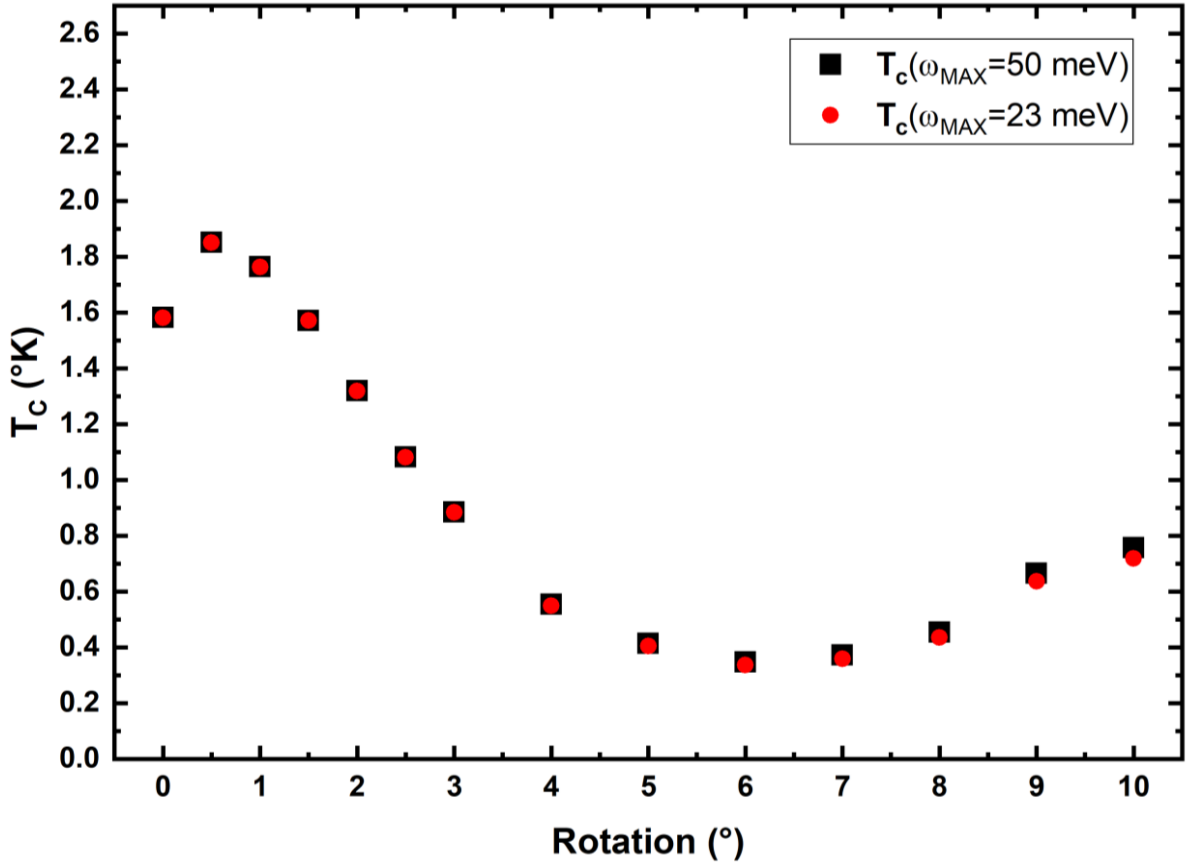

**Figure 8.** Superconducting critical temperatures for twisted bismuth bilayers.

## 4. Conclusion

Twisted bismuth bilayers have shown several interesting properties, such as the presence of sharp peaks near the Fermi level, due to border effects of the nanolayers, as well as superconductivity for all angles studied, with a maximum of 1.85 K for a rotation of 0.5 ° between layers.
The superconductivity seems to be of the conventional type, and the superconducting transition temperature appears to be electronically driven since there is an evident similarity between the behavior of the electronic density of states and that observed for the superconducting transition temperatures. Also, calculating the Debye temperature with two different frequency cutoffs have a negligible impact in the superconducting transition temperatures.

## Data availability statement

The datasets used and analyzed during the current study are available from the corresponding author on reasonable request.

## Acknowledgements

I.R. thanks CONAHCYT, for his postdoctoral fellowship. D.H.-R. acknowledges CONAHCYT for supporting his Ph.D. studies. A.A.V., R.M.V. and A.V. thank DGAPA-UNAM (PAPIIT) for continued financial support to carry out research projects under Grants No. IN116520 and No. IN118223. María Teresa Vázquez and Oralia Jiménez provided the information requested. Alberto Lopez[†] and Alejandro Pompa assisted with the technical support and maintenance of the computing unit at IIM-UNAM. Simulations were partially carried at the Computing Center of DGTIC-UNAM through project LANCAD-UNAM-DGTIC-131.

## Author contributions

I.R. and A.A.V. conceived this research and designed it with the participation of R.M.V., A.V., D.H.-R. All the simulations were done by I.R. All authors discussed and analyzed the results. I.R. and A.A.V. wrote the first draft and the other authors enriched the manuscript. All authors gave their consent for the publication of this manuscript.

## Competing Interests Statement

The authors declare no conflict of interest in this work.